# Structural relaxation and Jahn-Teller distortion of LaMnO$_3$ (001) surface

F.L. Tang [a*], M. Huang [b], W.J. Lu [a], W.Y. Yu [a]

[a] *State Key Laboratory of Gansu Advanced Non-ferrous Metal Materials, Department of Materials Science and Engineering, Lanzhou University of Technology, 730050, People's Republic of China*

[b] *Department of Physics, The University of Texas at Dallas, Richardson TX 75080, USA*

**ABSTRACT**

We studied in detail the structural relaxation and Jahn-Teller distortion in LaMnO$_3$ (001) surface of the orthorhombic phase by means of classical atomistic simulation. It is found that MnO$_2$-terminated surface is more energetically favorable than LaO-terminated surface by 0.34 eV. The standard deviation of Mn-O bond lengths of MnO$_6$ octahedra and Jahn-Teller distortion oscillate in LaMnO$_3$ (001) surface. Our simulated atomic displacements in the surface are compared with some *ab initio* studies.

[*]Corresponding author. Email address: tfl03@mails.tsinghua.edu.cn (F.L. Tang).

# 1. Introduction

The discovery of colossal magnetoresistance in rare-earth manganites, $La_{1-x}A_xMnO_3$ ($A$ = Ca, Sr, and Ba) with perovskite structure, has attracted much attention for their rich display of interesting basic physics problems and possible applications [1-3]. Their thin films play a fundamental role in many technological applications such as magnetoresistive devices, microelectronics, catalysis and cathode materials, where surface structure and its quality are of primary importance [4-7].

At high temperature ($T \geq 750$ K), $LaMnO_3$ has simple cubic perovskite structure. Based on this structure, surface energy, relaxation/reconstruction and polarization of $LaMnO_3$ were investigated [8-15]. A (1 × 2) $LaMnO_3$ (110) surface reconstruction was provided in four possible O-terminated surfaces with classical shell model calculation. The surface energy ($E_s$ = 1.45 J m$^{-2}$) of this (1 × 2) reconstructed surface is saturated only when six to eight near-surface atomic planes (including both Mn and La planes) are relaxed with considerable dipole moments perpendicular to the surface. Other three types of surfaces have larger $E_s$ = 1.5-5.2 J m$^{-2}$ [8]. Akhtar *et al.* studied $LaMnO_3$ (110) and (001) surfaces with Mn- and La- termination. Mn-terminated surfaces were found to be more stable. They also proposed that the surface structure is controlled by a subtle interplay between the different cation ionic strengths. In the case of the $LaMnO_3$ (110) surface, one of the Mn-terminated configurations was initially reconstructed to reduce the surface energy by removing a partial layer of oxygen [10]. Density-functional and plane-wave calculations showed that the antiferromagnetic (001) surfaces have lower surface energies than the ferromagnetic

(001) surfaces. Both the (001) and (110) surfaces reveal considerable atomic relaxations, up to the fourth plane from the surface, which reduce the surface energy by about a factor of 2. The calculated effective atomic charges and the electron density also indicated a considerable reduction of the Mn and O atom ionicity on the surface [10,11]. This classical shell model calculation method [8] had also been used to study $SrTiO_3$ (001) surface energies and surface relaxation [14,15]. Mastrikov *et al.* studied in detail the atomic/electronic structure [16] and thermodynamic properties [17] of the $LaMnO_3$ surfaces, in both cubic and orthorhombic phases. They found considerable electronic density redistribution near $LaMnO_3$ surface which could affect atomic and molecular adsorption. Their calculated effective atomic charges weakly depend on the magnetic structure and slab stoichiometry. The surface structure of Sr/Ca-doped $LaMnO_3$ was studied in some experiments [18-20]. Coaxial impact-collision ion scattering spectroscopy revealed that the (001) surface of *c*-axis oriented $La_{0.7}Sr_{0.3}MnO_3$ thin films is predominantly terminated with $MnO_2$ plane [20].

At low temperature ($T < 750$ K), $LaMnO_3$ has orthorhombic lattice structure [16,21], in which Jahn-Teller (JT) effect plays an important role by affecting lattice distortion and material properties (for example, charge/orbital ordering [22] and Curie temperature $T_C$ [1-3,23-26]). JT distortion in $LaMnO_3$ can lead its Fermi surface instability with decreasing temperature [27]. Zhao *et al.* [28] proposed a formula $T_C \propto W_{eff} \propto W \exp(-\gamma E_{JT}/h\omega)$ to qualitatively explain the isotope effect on $T_C$ with bandwidth $W$ and JT energy $E_{JT}$. Although the atomic/electronic structure and thermodynamic properties of the $LaMnO_3$ surfaces were well investigated in both cubic and orthorhombic phases in previous studies [7-14,16,17], the JT distortion in LaMnO$_3$ surfaces was not yet included in these studies. Since the JT distortion is important in orthorhombic-LaMnO$_3$, it would be useful to study the JT distortion in LaMnO$_3$ surfaces to further understand the surface properties. In this paper, we mainly used an atomistic simulation technique to study the details of the surface structure of orthorhombic LaMnO$_3$ with an emphasis on JT distortion.

## 2. Computational methods and surface models

The crystal structure of a bulk material at a given temperature and pressure can be predicted by minimizing its free energy. Our approach is to adjust the cell volume and atomic positions until the net pressure or stress is zero. During the iterative procedure, a constant volume energy minimization is performed. This technique has been used for simulation of many kinds of materials [23-26,29-32], especially for the Jahn-Teller distortion/energy in doped LaMnO$_3$ [23-26]. Details of this technique are available in [32].

The simulation technique is based on the widely used successful shell model [33] generalization of the Born model of a solid. With this model, the lattice energy $E$ can be expressed as

$$E = \frac{1}{2} \sum_{i,j} [\frac{q_i q_j}{r_{ij}} + V(r_{ij})], \qquad (1)$$

where the first item is Coulombic energy introduced by long range interactions of effective charges, and the second item is the short range interactions. Short-range interaction is represented by a Buckingham potential:

$$V(r) = A\exp(-r/\rho) - Cr^{-6}, \tag{2}$$

where $A$, $\rho$, and $C$ are fitting potential parameters. In order to describe the polarization of an individual ion and its dependence on local atomic environment, every ion is treated by the core-shell model. The interaction between the core and shell of any ion is considered as a harmonic spring [33].

The initial lattice structure we started for studying the $LaMnO_3$ surface is the crystallographic unit cell of $LaMnO_3$ (in Fig. 1), which has four $La^{3+}$ ions, four $Mn^{3+}$ ions, four O1 ions, and eight O2 ions (we denoted the $O^{2-}$ ions along the $c$ axis as O1, the $O^{2-}$ ions on the $a$-$b$ plane as O2 in $MnO_6$ octahedra). The available potential parameters for $LaMnO_3$ were obtained at 0 K by an empirical fitting method in [23]. We used different potential parameters for bonds of Mn-O1, Mn-O2, La-O1, and La-O2 so that the potentials can describe the directions of orbits of $Mn^{3+}$ $d$ electrons and Jahn-Teller effects. This potential can well reproduce the experimental crystal structure of bulk $LaMnO_3$. In [23], we had carefully investigated the pressure and temperature effect to further test these potential parameters. The calculated compressibility and thermal behavior are in good agreement with experimental results. In this work, we used these potential parameters [23-26] to study $LaMnO_3$ (001) surfaces.

To obtain a suitable surface slab model (Fig. 1) and make the calculations most efficient, the unit cell of $LaMnO_3$ is extended to two times along the $a$, $b$ axis directions and six times along the $c$ axis direction ($11.53 \times 11.08 \times 46.03$ Å$^3$). There are 480 ions (96 $La^{3+}$, 96 $Mn^{3+}$, and 288 $O^{2-}$ ions) in the surface slab model. The surface slab has planar two-dimension periodic boundary conditions parallel to the surface, as shown in Fig. 1. Along $c$ axis direction, there are twelve $MnO_2$ planes and twelve LaO planes. The slab is split into two regions (I and II). Above region I, there is a semi-infinite vacuum. During the simulation, the atoms of the region I structural units are relaxed explicitly until there is zero force on each of them, whilst those in region II are kept fixed to reproduce the potential of the bulk lattice on region I. The lattice parameters $a$ and $b$ of the slab are kept fixed during the simulation, thus the surface energy $E_s$ can be calculated [32] as

$$E_s = \frac{E_{slab} - mE_{bulk}}{A}, \tag{3}$$

where $E_{slab}$ is the total energy of the two-dimensional slab with $m$ $LaMnO_3$ formula units, $E_{bulk}$ is the total energy per unit of the $LaMnO_3$ bulk and $A$ is the surface area of the slab.

Charge redistribution is a general phenomenon at some oxides' polarized surface [11,16,17]. Our atomic simulation cannot describe this phenomenon very well because fixed charge states (+3 for La and Mn, -2 for O) were used in classical potential. Therefore, we performed density functional theory (DFT) calculations to $LaMnO_3$ (001) surface to find the charge difference of ions located in the bulk and those located at the surfaces. The DFT calculations are based on plane waves, generalized gradient corrected Perdew-Burke-Ernzerhof (PBE) approximation for the exchange and correlation functional, ultrasoft pseudopotentials and periodic supercells by using Quantum-ESPRESSO computer package [34-37]. The cores of Mn and La atoms were represented by small cores with atomic configurations of $3s^23p^63d^54s^2$ (15 valence electrons) for Mn and $5s^2 5p^6 5d^1 6s^2$ (11 valence electrons) for La. The atomic configuration for O atom is $2s^2 2p^4$ (6 valence electrons). The plane wave energy cutoff was set to be 30 Ry. The (001) surface was modeled with a slab consisting six $MnO_2$ planes and six LaO planes and a vacuum thickness of 12 Å. The atomic positions of $LaMnO_3$ (001) surface were relaxed according to the Hellmann–Feynman forces until the maximum atomic force was less than 0.02 eV/Å. By applying the charge redistribution data obtained from DFT study, we modified our classical charge states for (001) surface and optimized it again.

## 3. Results and discussion

### 3.1. Surface energy

$MnO_2$-/LaO-terminated surface configurations had been simulated. The total energy and surface energy of one surface configuration decrease as the number of the relaxed planes increases. Kotomin *et al.* pointed out that the $LaMnO_3$ (110) surface energy saturates when 3-4 Mn-O layers and 3-4 La-O layers were relaxed [7,8,16]. Twelve $MnO_2$ planes and twelve LaO planes in the whole slab were used in our atomistic simulation with upper half relaxed (region I) and lower half fixed (region II). Therefore, the models we used are large enough for the energy convergence.

The surface energy of $MnO_2$-/LaO-terminated surface is illustrated in Table 1. For $MnO_2$-/LaO-terminated surface, the calculated surface energy is 2.84/4.77 J m$^{-2}$ when the fixed bulk charge states (+3 for La and Mn, -2 for O) were applied to the whole slab. These values for surface energy are larger than those obtained from *ab initio* calculations shown in Table 1 [10,11,16]. The difference among them may arise from surface charge redistribution. In order to investigate the charge difference of ions located in the bulk and at the surface, we performed DFT calculations on the LaMnO$_3$ (001) surface. The Lowdin charge of an atom was obtained by projecting the plane-wave electronic density over virtual atomic *s, p* and *d*-orbitals. The calculated charges, relative charges and modified charges used in classical simulations for bulk atoms and (001) surface atoms were shown in Table 2. The relative charge is defined as the charge difference between the charge of an atom (La, Mn or O) at the surface and the charge for it in bulk. The calculated charges for most of the surface atoms are smaller than the relevant formal charges of +3, +3 and -2 on La, Mn, and O atoms, respectively, which is consistent with an early study [11]. However, our calculated charges are considerably smaller than the Bader charges [11]. The relative charge was introduced and applied to modify the charges of atoms in LaMnO$_3$ (001) surface, which would be used in our classical simulations. Take the case of a Mn atom in the first MnO$_2$ plane, Lowdin population analysis showed that a Mn atom in the bulk has 14.6766 valence electrons (charge is +0.3234) around it, and a Mn atom at the first MnO$_2$ plane has 14.7651 valence electrons (charge is +0.2349). The relative charge is -0.0885. In our classical simulations, the charge for bulk Mn is fixed as +3. We modified the charge state of Mn in the first MnO$_2$ plane in our classical potential by keeping the relative charge of Mn the same as that of our DFT results: Mn ion in the bulk has +3 charge state otherwise +2.9115 at the first MnO$_2$ plane (Fig. 1). Analogically, we obtained the relative charges and modified charges of other atoms located in LaMnO$_3$ (001) surface (from the first to the third MnO$_2$ and LaO plane) as shown in Table 2. The relative charges and modified charges for other planes are not shown for saving space. We used the modified charge states to optimize the surface structure again and obtained the surface energy: 1.05/1.39 J m$^{-2}$ for MnO$_2$-/LaO-terminated surface, qualitatively agreeing with the data given by other studies [10,11,16] as shown in Table 1 (0.74-0.94/1.21-1.79 J m$^{-2}$).

From Table 1, it is found that the relaxed surface with modified charges has more accurate surface energy than that from classical fixed charges. This indicates that there is a subtle difference between the atoms located in bulk positions and atoms located at the surface. Our *ab initio* calculation suggests that the surface charge redistribution plays an important role in determining surface energy and surface relaxation (as shown below). The bulk charge states can not be directly used for surface simulation, at least, for orthorhombic LaMnO$_3$ (001) surface simulation.

In Table 1, MnO$_2$-terminated surface has smaller surface energies than those of LaO-terminated surface: not only for relaxed surface with bulk charge states but also relaxed surface with modified charge states. So we predict that the MnO$_2$-terminated LaMnO$_3$ (001) surface is energetically favorable, which is in agreement with other calculations [10] and experimental observations [20].

*3.2. Surface structural relaxation*

The relaxed MnO$_2$-terminated LaMnO$_3$ (001) surface (Table 1, $E_s$ = 1.05 J m$^{-2}$) was selected to investigate the surface structural relaxation. We provide the atomic displacements of a series of MnO$_6$ octahedra and La ions (Fig. 2a) with approximately same *x* and *y* coordinates in the surface slab model. Three-dimension atomic displacements in the upper six $MnO_6$ octahedra and La ions are shown in Figs. 2b-2d. Fig. 2b shows the atomic displacements along *x* axis. It is found that La ions located in the odd planes (1, 3 and 5 planes in Fig. 1) shift along *x* axis direction 2-4% (in percent of the distance between two adjacent $MnO_2$ planes in the bulk $a_0 = 3.8315$ Å). Otherwise, La ions in the even planes (2, 4 and 6 planes in Fig. 1) displace along –*x* direction about 4-8%. Here we denote –*x*, –*y* and –*z* as the opposite direction of *x*, *y* and *z*, respectively. O2 ions in the first, the third or the fifth $MnO_6$ octahedron shift along the same direction 5-8%. Otherwise, in the second, the fourth or the sixth $MnO_6$ octahedron, two O2 ions shift along *x* about 7%, and the other two O2 ions shift along –*x* about 6%. The first MnO plane has the smallest displacement along –*x*. Atomic displacements in other MnO planes decrease as the number labeled for plane increases.

Figure 2c shows the atomic displacements along *y* axis. It is found that O2 ions on the first plane have largest displacements (the average value is about 11%). Ions on other planes displace along *y* axis slightly (< 4%). Mn ions on the first and the third $MnO_2$ plane shift with 5% displacement but other Mn ions have displacements less than 2%. In the same $MnO_6$ octahedron, two O2 ions shift along one direction and the other two O2 ions will shift along the opposite direction.

Figure 2d shows the atomic displacements along *z* axis. It is clear that all Mn ions shift outwards about 1-2%. On the first $MnO_2$ plane, two O2 ions shift along –*z* about 8%, and the other two O2 ions shift along *z* about 3%. We noted that four O2 ions on the uppermost plane tend to form a flat face parallel to *x-y* plane, but the four O2 ions in other $MnO_6$ octahedra form a slope flat face relative to *x-y* plane. On other $MnO_2$ planes, O2 ions shift along the same direction –*z* 4-5%. It is also found that all La ions shift outwards, and the displacement decreases from 10% to 2% as the plane number increases from one to six. All the O1 ions in LaO planes shift with small displacements (< 2%).

Our simulate results can be compared with *ab initio* calculation data. Evarestov *et al.* [11] found that the first Mn ion on the outermost surface shifts outward from the surface slab center 0.45% (average value from Table IV in [11], in percent of the lattice parameter of the cubic $LaMnO_3$, $a_0 = 3.95$ Å). But the second, the third and the fourth Mn ions shift towards the slab center with the displacement 0.7%, 0.7% and 0.54%, respectively. The first, the second and the third La ions shift outwards with the displacement 7.88%, 4.54% and 4.7%, respectively. We have the same conclusions on La ion displacements in LaMnO3 (001) surface as those in *ab initio* simulation [11]: the same displacement direction (outwards) and approximately equal shifting distance (4-8%). But we have different suggestions for the Mn ion displacements on the second to the fourth Mn planes. Mastrikov *et al.* observed that the largest atomistic displacement took place along *z* axis in $LaMnO_3$ (001) surfaces [16]. We found that La ions have the largest displacement along *z* axis. But O ions have larger displacements along *x* axis and along *z* axis (Fig. 1b-d). Just as pointed in Ref. 16, we also found that LaO planes are split into La plane and O plane (La ions shift outwards and O ions inwards). The calculated Mn-O distances along the *z* axis [17] showed that the upper parts of $MnO_6$ octahedra were regularly expanded whereas the lower parts compressed as compared to the perfect bulk $MnO_6$ octahedron. We also found that the upper Mn-O bond lengths of $MnO_6$ octahedra along $z$ axis range from 2.000 Å to 2.029 Å whereas the lower Mn-O bond lengths range from 1.910 Å to 1.928 Å (the corresponding bond length is 1.961 Å in the perfect bulk $MnO_6$ octahdera). Our classical simulation has the same conclusion on $MnO_6$ octahedral distortion along z axis as shown in [17].

From Fig. 2, we concluded that (1) La ions shift mainly along $z$ direction; (2) Mn ions shift mainly in $x$-$y$ plane; (3) All Mn ions and La ions shift outwards; (4) O2 ions have larger displacements than other ions along three directions; (5) O1 ions have the smallest displacement; (6) From the second planes to the seventh planes, atomic displacements along $y$ axis are much smaller than those along $x$ and $z$ axes. These local structure deviations may lead to electronic structure inhomogeneity, which can affect magnetic and electronic properties of this compound, especially the surface properties. In hole-doped manganites, Mn valency can be changed from four in the bulk to three at the (001) subsurface [12].

Since O2 ions have larger displacement as mentioned above, we compare their atomic positions with their corresponding bulk positions in $x$-$y$ plane (Fig. 3). Fig. 3a shows the atomic displacements of the first $MnO_2$ plane compared with their corresponding bulk positions. It is found that the four O2 ions around Mn ion at A site rotate counter-clockwise, but the four O2 ions around Mn ion at B site rotate clockwise. At A and B sites, the longer Mn-O bond lengths (along [110] at A site;

along [$\bar{1}$10] at B site) are elongated; the short Mn-O bond lengths (along [$\bar{1}$10] at A site; along [110] at B site) are shortened. The average Mn-O-Mn bond angle increases from 154.6° in the bulk to 175.2° at the first MnO$_2$ plane. Fig. 3b shows displacements of atoms in the second MnO$_2$ plane compared with their corresponding bulk positions. The four O2 ions around Mn ion at C site rotate counter-clockwise, but the four O2 ions around Mn ion at D site rotate clockwise. Different from the cases of the first MnO$_2$ plane, at C and D sites, the longer Mn-O bond lengths (along [110] at C site; along [$\bar{1}$10] at D site) become shortened; the short Mn-O bond lengths (along [$\bar{1}$10] at C site; along [110] at D site) are elongated. The averaged Mn-O-Mn bond angle changes from 154.6° in the bulk to 165.7° at the second MnO$_2$ plane. Fig. 3c shows displacements of atoms in the third MnO$_2$ plane compared with their corresponding bulk positions. It seems that all O2 ions have a parallel movement (9%) along [100] direction. Similar to the cases for second MnO$_2$ plane, at E or F site, the longer Mn-O bond length (along [110] at E site; along [$\bar{1}$10] at F site) is shortened; the short Mn-O bond length (along [$\bar{1}$10] at E site; along [110] at F site) is elongated. The averaged Mn-O-Mn bond angle decreases from 154.6° in the bulk to 144.8° at the third MnO$_2$ plane. Fig. 3d shows the displacement of atoms on the fourth MnO$_2$ plane. O2 atoms in the fourth MnO$_2$ plane show similar behaviors as those in the second MnO$_2$ plane, but smaller displacement resulting in the averaged bond angle (156.5°) being close to that in perfect bulk (154.6°).

*3.3. Octahedral distortion and Jahn-Teller distortion*

The displacements of Mn-O bonds are different for different MnO$_6$ octahedra in the surface slab. Therefore we studied the standard deviation $SD_{Mn-O}$ of Mn-O bond lengths in $MnO_6$ octahedra which was shown in Fig. 4a using the formula $SD_{Mn-O} = \sqrt{\sum_{i=1}^{n}(l_i - \bar{l})^2/n}$, where $l_i$ is the $i$th Mn-O bond length and $\bar{l}$ is the average value of all of the $n$ bond lengths ($n$ = 5 for the first octahedron of Mn-terminated surface and $n$ = 6 for those in other octahedra). The surface $SD_{Mn-O}$ was found to be larger than the bulk value for both $MnO_2$-terminated surface and LaO-terminated surface. As the number of the octahedron increases from one to seven or more, $SD_{Mn-O}$ in $LaMnO_3$ (001) surface decreases from 0.4 Å ($MnO_2$-terminated) or 0.26 Å (LaO-terminated) to 0.1 Å, indicating that the distortion of $MnO_6$ octahedra in the surface is larger than that in the bulk. We also found that $SD_{Mn-O}$ oscillates in the surface for $MnO_2$-/LaO-terminated surface. $SD_{Mn-O}$ reaches a local maximum on the third/fifth $MnO_2$ plane and a local minimum on the second/fifth $MnO_2$ plane.

Taking into account lattice distortions only, the cooperative Jahn-Teller distortion $Q_{JT}$ can be presented using the following formula [38]:

$$Q_{JT} = \sqrt{(Q_{2u} + Q_{2s})^2 + (Q_{3u} + Q_{3s})^2} \\ + \sqrt{(Q_{2u} - Q_{2s})^2 + (Q_{3u} - Q_{3s})^2}], \quad (4)$$

where $Q_{2s}$ and $Q_{3s}$ can be easily calculated from the coordinates of six O atoms in one $MnO_6$ octahedron, and $Q_{2u}$, $Q_{3u}$ from lattice parameters. The $Q_{JT}$ on a specific $MnO_6$-plane is calculated by averaging the $Q$ of every octahedron considered. In equation (4),

$$Q_{2u} = \frac{a_0}{\sqrt{2}}(e_{xx} - e_{yy}), \quad (5)$$

$$Q_{3u} = \frac{a_0}{\sqrt{6}}(2e_{zz} - e_{xx} - e_{yy}), \tag{6}$$

$$Q_{2s} = \frac{a_0}{\sqrt{2}}(v_{sx} - v_{sy}), \tag{7}$$

$$Q_{3s} = \frac{a_0}{\sqrt{6}}(2v_{sz} - v_{sx} - v_{sy}), \tag{8}$$

where $a_0$ is the lattice parameter of the ideal cubic perovskite structure. In this work, we set $a_0 = b/2 = 3.8315$ Å. In equations (5) and (6), $e_{jj}$ is the diagonal component of the conventional strain tensor referred to the ideal cubic perovskite lattice. In equations (7) and (8) $v_i^a = u_i^a - u_{i-a}^a$, where $u_i^a$ and $u_{i-a}^a$ are displacements of the two O ions in the $MnO_6$ octahedron along the $a$ direction. The details about the formulas can be found in references [1] and [38].

Using lattice transitions and octahedral distortions, we calculated Jahn-Teller distortion $Q_{JT}$ of $MnO_6$ octaheda shown in Fig. 4b. The calculated $Q_{JT}$ increases from 1.6 Å ($MnO_2$-terminated surface) or 1.0 Å (LaO-terminated surface) to 2.5 Å as $MnO_6$ octahedron number increases from one (LaO-terminated surface) or two ($MnO_2$-terminated surface, the first $MnO_6$ octahedron is not completed as it just has five Mn-O bonds) to seven or more. Similar to $SD_{Mn-O}$ of Mn-O bond lengths, $Q_{JT}$ was found to show oscillating behavior in the $LaMnO_3$ (001) surface.

The standard deviation $SD_{Mn-O}$ is computed from Mn-O bond lengths whereas $Q_{JT}$ from lattice transitions and octahedral distortions. In addition, it is found that the octahedral distortions $v_{sx}$, $v_{sy}$, and $v_{sz}$ counteract with each other in the surfaces. This maybe the reason why $SD_{Mn-O}$ is larger whereas $Q_{JT}$ is smaller in the surfaces compared with their bulk values.

## 4. Conclusion

Using atomistic simulation and classical potentials modified by *ab initio* charge redistribution, we studied the surface relaxation and Jahn-Teller distortion of orthorhombic LaMnO$_3$ (001) surfaces. Specific conclusions are as follows:

(1) MnO$_2$-terminated surface (the surface energy $E_s$ = 1.05 J m$^{-2}$) is more energetically favorable than LaO-terminated surface ($E_s$ = 1.39 J m$^{-2}$).

(2) For MnO$_2$-terminated surface, oxygen ions on MnO$_2$ planes have the largest displacement, whereas oxygen ions on LaO planes have the smallest displacement. La and Mn ions shift outwards.

(3) The standard deviation $SD_{Mn-O}$ among Mn-O bond lengths of surface MnO$_6$ octahedra (0.2-0.4 Å) is larger than that in the bulk (0.1 Å), but surface Jahn-Teller distortion $Q_{JT}$ (1.0-2.0 Å) is smaller than the bulk value (2.5 Å).

(4) Both $SD_{Mn-O}$ and $Q_{JT}$ oscillate in LaMnO$_3$ (001) surface.

**Acknowledgment**

The authors would like to thank financial support by LUT Research Development Funding (BS01200905) and National Science Foundation Plan Program. This work is mainly performed in Gansu Province Supercomputer Center with Y.L. Shen's technical help.

**Table 1** Surface energies (J m$^{-2}$) of LaMnO$_3$ (001) surfaces.

|  | MnO$_2$-terminated | | LaO-terminated | |
|---|---|---|---|---|
| This work | 2.84$^a$ | 1.05$^b$ | 4.77$^a$ | 1.39$^b$ |
| Ref. [10] | 0.94 | | 1.21 | |
| Ref. [11] | 0.74 | | 1.79 | |
| Ref. [16] | | 0.85 | | |

$^a$Relaxed surface with fixed bulk charges (+3 for La and Mn, -2 for O);
$^b$Relaxed surface with modified charges (Table 2);
Ref. [10]: from Table 3 in [10];
Ref. [11]: from Table III in [11], the surface slab had 10 planes, the unit eV/$a_0$ ($a_0$ = 3.95 Å) is changed into J m$^{-2}$;
Ref. [16]: from Table 2 in [16], the surface slab had 8 planes, the unit eV/$a_0$ ($a_0$ = 3.95 Å) is changed into J m$^{-2}$.

**Table 2** The calculated charges, relative charge and modified charges used in classical simulations for bulk LaMnO$_3$ atoms and (001) surface atoms. The relative charge is defined as the charge difference between the charge of an atom (La, Mn or O) at the surface and the charge for it in bulk. The first to the third MnO$_2$ or LaO planes can be found in Fig. 1.

|  |  | Electrons ($e$) (DFT results) | Charge (DFT results) | Relative charge (DFT results) | Modified charge |
|---|---|---|---|---|---|
| Bulk | La | 10.0037 | 0.9963 | 0.0 | +3 |
|  | O1 | 6.3955 | -0.3955 | 0.0 | -2 |
|  | Mn | 14.6766 | 0.3234 | 0.0 | +3 |
|  | O2 | 6.4057 | -0.4057 | 0.0 | -2 |
| 1$^{st}$ MnO$_2$ plane | Mn | 14.7651 | 0.2349 | -0.0885 | +2.9115 |
|  | O2 | 6.4435 | -0.4435 | -0.0378 | -2.0378 |
| 1$^{st}$ LaO plane | La | 10.0178 | 0.9822 | -0.0141 | +2.9859 |
|  | O1 | 6.3521 | -0.3521 | 0.0434 | -1.9566 |
| 2$^{nd}$ MnO$_2$ plane | Mn | 14.8130 | 0.1870 | -0.1364 | +2.8636 |
|  | O2 | 6.4362 | -0.4362 | -0.0305 | -2.0305 |
| 2$^{nd}$ LaO plane | La | 10.0083 | 0.9917 | -0.0046 | +2.9954 |
|  | O1 | 6.5138 | -0.5138 | -0.1183 | -2.1183 |
| 3$^{rd}$ MnO$_2$ plane | Mn | 14.6456 | 0.3544 | 0.0310 | +3.0310 |
|  | O2 | 6.4032 | -0.4032 | 0.0025 | -1.9975 |
| 3$^{rd}$ LaO plane | La | 9.9976 | 1.0024 | 0.0061 | +3.0061 |
|  | O1 | 6.3343 | -0.3343 | 0.0612 | -1.9388 |

**Figure captions**

Fig. 1. (Color online) LaMnO$_3$ (001) surface slab model. Dashed rectangle contains the crystallographic unit cell of LaMnO$_3$ from which the surface slab is built. This surface slab model is also extended to two times along the *a*, *b* axis directions, not shown for simplicity.

Fig. 2. (Color online) MnO$_6$ octahedra and La ions (a) in MnO$_2$-terminated LaMnO$_3$ (001) surface (Fig. 1) and their displacements *dx* (b), *dy* (c) and *dz* (d) (in percent of the distance between two adjacent MnO$_2$ planes $a_0$ = 3.8315 Å) along *x*, *y*, *z* directions, respectively. Positive displacement indicates that the ion shifts along the axis direction; negative value stands for displacement along the opposite direction of the axis.

Fig. 3. (Color online) Comparison of the first (a), the second (b), the third (c) and the fourth (d) MnO$_2$ surface planes with their corresponding bulk positions. The surface ions cover the bulk ions.

Fig. 4. (Color online) Standard deviation of Mn-O bond lengths (a) and Jahn-Teller distortion (b) of MnO$_6$ octahedra in MnO$_2$- and LaO-terminated LaMnO$_3$ (001) surfaces.

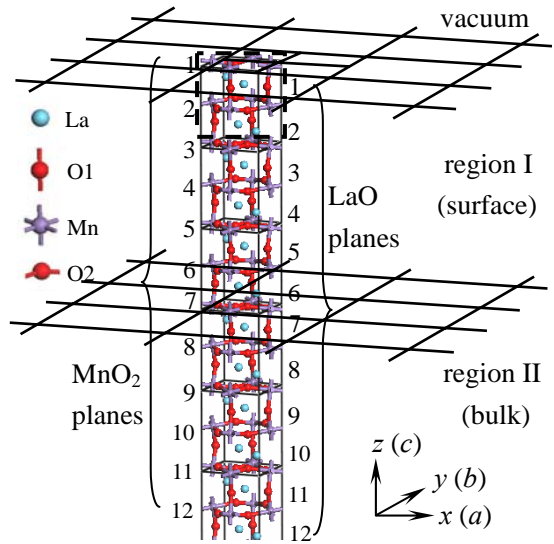

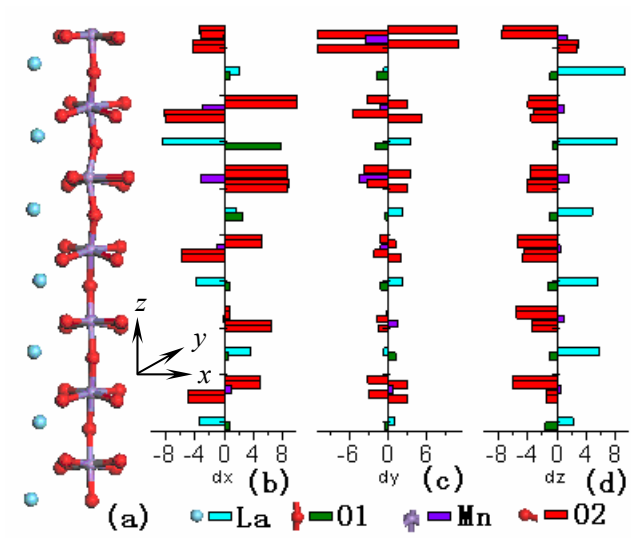

Fig. 2

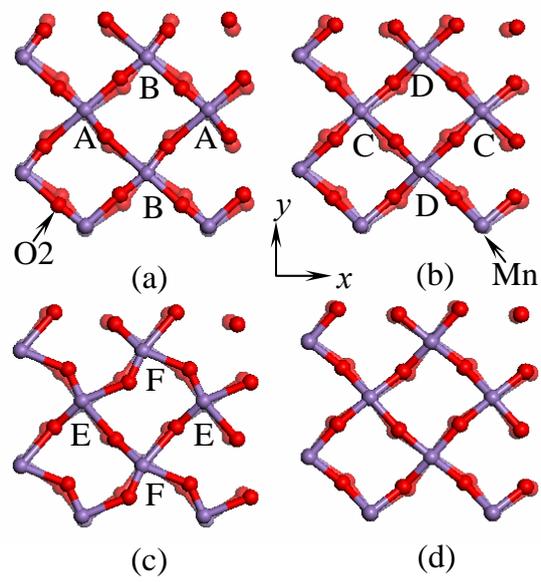

Fig. 3

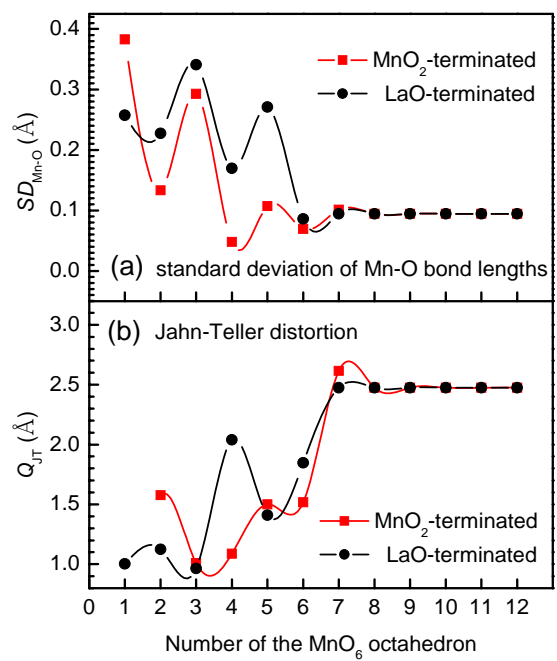